\documentclass{aa} 

\usepackage{graphicx}
\bibpunct{(}{)}{;}{a}{}{,} 
\usepackage{natbib}
\usepackage{amsmath,multirow}
\usepackage{txfonts}
\usepackage[colorlinks=true,urlcolor=blue,citecolor=blue,linkcolor=blue]{hyperref}

\usepackage{url,lineno,microtype,subcaption}

\begin{document}

   \title{Extreme-ultraviolet transient brightenings in the quiet-Sun corona}
   
   	\subtitle{Closest-perihelion observations with Solar Orbiter/EUI}

    \author{Nancy Narang  \inst{1}
   	\and
   	Cis Verbeeck \inst{1}
   	\and
   Marilena Mierla  \inst{1,7}
   \and
   David Berghmans  \inst{1}
   \and
    Fr\'ed\'eric Auch\`ere  \inst{2}
    \and
    Sergei Shestov  \inst{1,3}
    \and
    V\'eronique Delouille  \inst{1}
    \and
     Lakshmi Pradeep Chitta  \inst{4}
    \and 
    Eric Priest  \inst{5}
    \and
    Daye Lim  \inst{6,1}
   \and
   Laurent R. Dolla \inst{1}
   \and
    Emil Kraaikamp  \inst{1}
    }

   \institute{Solar-Terrestrial Centre of Excellence -- SIDC, Royal Observatory of Belgium, Ringlaan -3- Av. Circulaire, 1180 Brussels, Belgium\\
	\email{nancy.narang@oma.be}\\
        \and
        Universit\'e Paris-Saclay, CNRS, Institut d'Astrophysique Spatiale, 91405 Orsay, France\\
        \and
        Centre Spatial de Li\`ege, Universit\'e de Li\`ege, Av. du Pr\'e-Aily B29, 4031 Angleur, Belgium\\
         \and
        Max-Planck-Institut f\"ur Sonnensystemforschung, Justus-von-Liebig-Weg 3, 37077 G\"ottingen, Germany \\
           \and Mathematics Institute, St Andrews University, KY16 9SS, St Andrews, UK \\
            \and
           Centre for mathematical Plasma Astrophysics (CmPA), Department of Mathematics, KU Leuven, 3001 Leuven, Belgium\\
           \and
         Institute of Geodynamics of the Romanian Academy, Bucharest, Romania
    }


 
  \abstract
   {The extreme-ultraviolet (EUV) brightenings identified by Solar Orbiter, commonly known as \textit{campfires}, are the smallest transient brightenings detected to date outside active regions in the solar corona.
   } 
    {To understand their possible contribution to quiet-Sun heating, we investigate the spatio-temporal distribution of a large ensemble of the finest-scale EUV transient brightenings observed by the Extreme Ultraviolet Imager (EUI) aboard Solar Orbiter.
    }
   {We perform a statistical analysis of the EUV brightenings by using quiet-Sun observations at the highest possible spatial resolution ever obtained by EUI. We use observations in the 17.4\,nm passband of the High Resolution EUV Imager (HRI\textsubscript{EUV}) of EUI acquired during the closest perihelia of Solar Orbiter in 2022 and 2023. Solar Orbiter being at a distance 0.293\,AU from the Sun, these observations have an exceptionally high image scale of 105\,km, recorded at a fast cadence of 3\,seconds. We use a wavelet-based automatic detection algorithm to detect and characterise the events of interest, and study their morphological and photometrical properties.
	}
   {We report the detection of smallest and shortest lived EUV brightenings to date in the quiet-Sun. The size and lifetime of the detected EUV brightenings appear power-law distributed down to a size of 0.01\,Mm$^{2}$ and a lifetime of 3\,seconds. In general their sizes lie in the range of 0.01\,Mm$^{2}$ to 50\,Mm$^{2}$, and their lifetimes vary between 3\,seconds and 40\,minutes. We find an increasingly high number of EUV brightenings at smaller spatial and temporal scales. We estimate that about 3600 EUV brightenings appear per second on the whole Sun. The HRI\textsubscript{EUV} brightenings thus represent the most prevalent, localised and finest-scale transient EUV brightenings in the quiet regions of the solar corona.
   	}
   {By using observations from EUI/HRI\textsubscript{EUV} at the highest possible achievable spatial resolution with the fastest cadence ever attained for quiet-Sun EUV observations, we detect the smallest and shortest-lived EUV brightenings to date. Future studies that can provide estimates of the thermal energy content of the smallest-scale EUV brighteninigs will help to gain better insights into their role in the coronal heating.}


   \maketitle
   \nolinenumbers
%

\section{Introduction}

\label{sect1}

 The ubiquitous presence of localised transient brightenings throughout the solar atmosphere is generally attributed to an impulsive release of energy by the process of magnetic reconnection \citep{2015Klimchuk, pontin2022}. The nanoflare heating model of the quiet solar corona proposed in \citet{1988Parker} and developed in \citet{priest2002} as the tectonics model involves the prevalence of transient events or bursts, driven by magnetic reconnection \citep{1972Parker, 1983aParker, 1983bParker} due to misaligned magnetic field strands, occurring continuously at small spatial and temporal scales in the solar atmosphere. This misalignment and structuring of the events in the corona is likely caused by the rapidly evolving small-scale magnetic elements on the solar surface. For instance, \citet{smitha2017} discovered an order of magnitude more photospheric flux cancellation events than previously known at the boundaries of many granules, which has led to the development of a flux cancellation model \citep{priest2018} for heating the chromosphere and corona and even accelerating the solar wind \citep{pontin24}.
 
 In the solar transition region and lower corona, a plethora of small-scale localised brightening events in the quiet-Sun, with typical length scales of a few Mm, has been documented in the literature \citep{2003Harrison, 2018Young}. Transition region explosive events (TREEs) are the most widely studied in this category \citep{1983Brueckner, 1989Dere, 1994Dere, 1997Innes, 2002Teriaca, 2004Teriaca}. TREEs are generally characterized by their broad and multi-component spectral profiles in typical transition region emission lines. Other classes of localised brightenings in the quiet-Sun detected from various EUV imaging instruments include, but are not limited to, Blinkers \citep{1997Harrison}, EUV transients \citep{1998Berghmans} and many other micro/nanoflare type transient events \citep{1998Krucker, 2000aAschwanden, 2000bAschwanden, 2000Parnell, 2002Benz, 2016Joulin, 2021Chitta, 2022Phurkart, 2024Belov}.  Despite the detections of a variety of EUV transients in the micro- to nano-flare energy range, their estimated energy flux rates have been reported to be insufficient to balance the coronal quiet-Sun energy losses \citep{1977Withbroe, 2016Aschwanden}. Given the powerlaw-like distribution of the sizes and lifetimes of transient events in the corona, the determination of these characteristics over finer scales has remained a reflection of the limitations of the instruments, in terms of spatial and temporal resolutions, rather than an actual characteristation of the events.

Recently, \citet{2021Berghmans} have reported the smallest EUV transient brightenings or bursts in the solar corona to date, identified by the High-Resolution EUV Imager (HRI\textsubscript{EUV}) of Extreme Utraviolet Imager (EUI, \citealt{2020Rochus}) onboard Solar Orbiter \citep{2020Muller}. Commonly known as \textit{campfires}, these EUV brightenings appear in the lower corona and above the locations of supergranular network boundaries. They have been reported to have sizes between 0.08\,Mm$^{2}$ and 5\,Mm$^{2}$ and durations between 10\,s and 200\,s by \citet{2021Berghmans}. Combining the observations of HRI\textsubscript{EUV} and AIA (Atmospheric Imaging Assembly, \citealt{2012Lemen}) onboard SDO (Solar Dynamics Observatory, \citealt{2012Pesnell}) and using stereoscopic techniques \citet{2021Zhukov} have found that these EUV brightenings are typically formed over the transition region to lower corona, \textit{i.e.,} at heights between 1\,Mm to 5\,Mm in the solar atmosphere. 

Recent magnetohydrodynamic simulations \citep{2021Chen, 2025Chen} and magnetic field extrapolations \citep{2022Barczynski} have indicated that magnetic reconnection in the transition region and lower corona is the likely mechanism for the generation of EUV brightenings observed by HRI\textsubscript{EUV}. By studying the photospheric magnetic field at the base of the EUV brightenings \citet{2021Panesar} and \citet{2022Kahil} have shown that many of the EUV brightenings observed by HRI\textsubscript{EUV} occur above the locations of small-scale bipolar magnetic elements, which is consistent with the flux-cancellation model \citep{priest2018,pontin24}. Furthermore, \citet{2024Nelson} have found that these EUV brightenings preferentially appear co-spatial with the strong ($>$\,20 G) photospheric network field, though they predominantly do not occur co-spatial with bipolar magnetic elements.

 Using EUV observations from the AIA \citet{2021Chitta} have also reported the presence of EUV transient brightenings or EUV bursts in the quiet-Sun. These EUV bursts were reported to have sizes in the range of 0.2\,Mm$^{2}$ to 10\,Mm$^{2}$ and lifetimes between 36\,s and 400\,s in \citet{2021Chitta}. The occurrence rate of these EUV bursts was found to be a factor of 100 less than that required to balance the quiet-Sun energy requirement. The EUV brightenings reported in \citet{2021Berghmans} have an occurrence rate of almost twice that of the EUV bursts reported in \citet{2021Chitta}. The higher spatial and temporal resolution of HRI\textsubscript{EUV}, and its higher radiometric sensitivity \citep{2023Gissot, Shestov25} in comparison to AIA can be considered as an important reason for the higher number of detections in the former case. The HRI\textsubscript{EUV} observations in \citet{2021Berghmans} have a one-pixel resolution of 200\,km and a cadence of 5\,s, while the AIA observations in \citet{2021Chitta} have those of 435\,km and 12\,s. Even with the high-resolution observations of HRI\textsubscript{EUV}, the number of EUV brightenings reported in \citet{2021Berghmans} still represents a deficit of at least 50 times the required rate to account for the quiet-Sun coronal energy requirements. This deficit can be improved by using much better spatially and temporally resolved observations to aid the detections of yet unresolved structures in the solar corona.
 
At the same time, there is also an active debate on the thermal properties of EUV brightenings observed by EUI along with their overall contribution to the coronal heating requirements. Based on imaging observations of EUI with AIA, \citet{2021Berghmans} found that the EUV brightenings reach the temperatures of 1\,MK, while \citet{2023Dolliou} have indicated that this may not be the case for all EUV brightenings. Additionally, by studying a sample of the EUV brightenings with UV/EUV spectroscopic observations,  \citet{Huang2023} and \citet{2024Dolliou} showed that many of the EUV brightenings may have temperatures below 1\,MK.

 As is the case for all observations, the smallest size and lifetime reported for the EUV brightenings observed by HRI\textsubscript{EUV} and the EUV bursts observed by AIA are also limited by the spatial and temporal resolution of the respective observations employed. Here, we present the HRI\textsubscript{EUV} observations taken during the closest perihelion of Solar Orbiter at about 0.293\,AU from the Sun. The closest proximity to the Sun that Solar Orbiter can achieve, these unique observations have the highest possible spatial resolution that can be achieved by HRI\textsubscript{EUV}. These observations have spatial and temporal resolutions of about two times better than those used in \citet{2021Berghmans} and the other EUV brightenings related studies using HRI\textsubscript{EUV} mentioned above, and four times better than the AIA observations used in \citet{2021Chitta}. The HRI\textsubscript{EUV} passband centered at the 17.4 nm has a peak temperature response close to 1\,MK. We take advantage of the unique EUV observations of the quiet-Sun corona with unprecedented resolution provided by HRI\textsubscript{EUV} at close proximity to the Sun to perform a statistical study of the properties of previously inaccessible small-scale EUV brightenings. As these brightenings are imprints of fine-scale magnetic reconnection, we explore their possible role in heating quiet regions of the solar corona. We use the term EUV brightenings for the transient small-scale EUV brightenings or EUV bursts observed by HRI\textsubscript{EUV} up to its resolution limit.




\section{Solar Orbiter/EUI perihelion observations}

\label{sect2}

 Solar Orbiter has a unique orbit \citep{2020Muller, 2020Zouganelis} and, by virtue of this, it can achieve the closest approach to the Sun (Solar Orbiter perihelion) that is only 0.293\,AU away from the Sun. \citet{2023Berghmans} has showcased the details of EUI observations for the Solar Orbiter's first close perihelion. In this study we use the quiet-Sun observations, at the Solar Orbiter perihelion (0.293\,AU), of HRI\textsubscript{EUV} in the 17.4\,nm passband taken on 2022-10-12 from 05:25:00\,UTC to 06:09:30\,UTC (dataset-1) and on 2023-10-06 from 10:00:00\,UTC to 11:00:00\,UTC (dataset-2). Both the datasets were obtained as the Solar Orbiter Observing Plan named R\_SMALL\_HRES\_HCAD\_RS-burst \citep{2020Zouganelis}.
 The HRI\textsubscript{EUV} images have a pixel scale of 0.492$^{\prime\prime}$ \citep{2023Gissot}, which for these observations correspond to approximately 105\,km on the solar surface. These observations has the highest possible spatial resolution that can be achieved by HRI\textsubscript{EUV}. They have an extremely high cadence of 3\,s, which is the best suitable fast cadence with which HRI\textsubscript{EUV} can obtain good quality quiet-Sun observations. These unique HRI\textsubscript{EUV} datasets thus constitute the best resolved observations, spatially and temporally, observations of the of the quiet Sun ever obtained in EUV passbands. The observations used here are the best of their kind, to date, to study the EUV brightenings occurring at the finest spatial and temporal scales. 
 
\begin{figure*}[htbp]
 	\centering
 	
 	\includegraphics[width=\hsize]{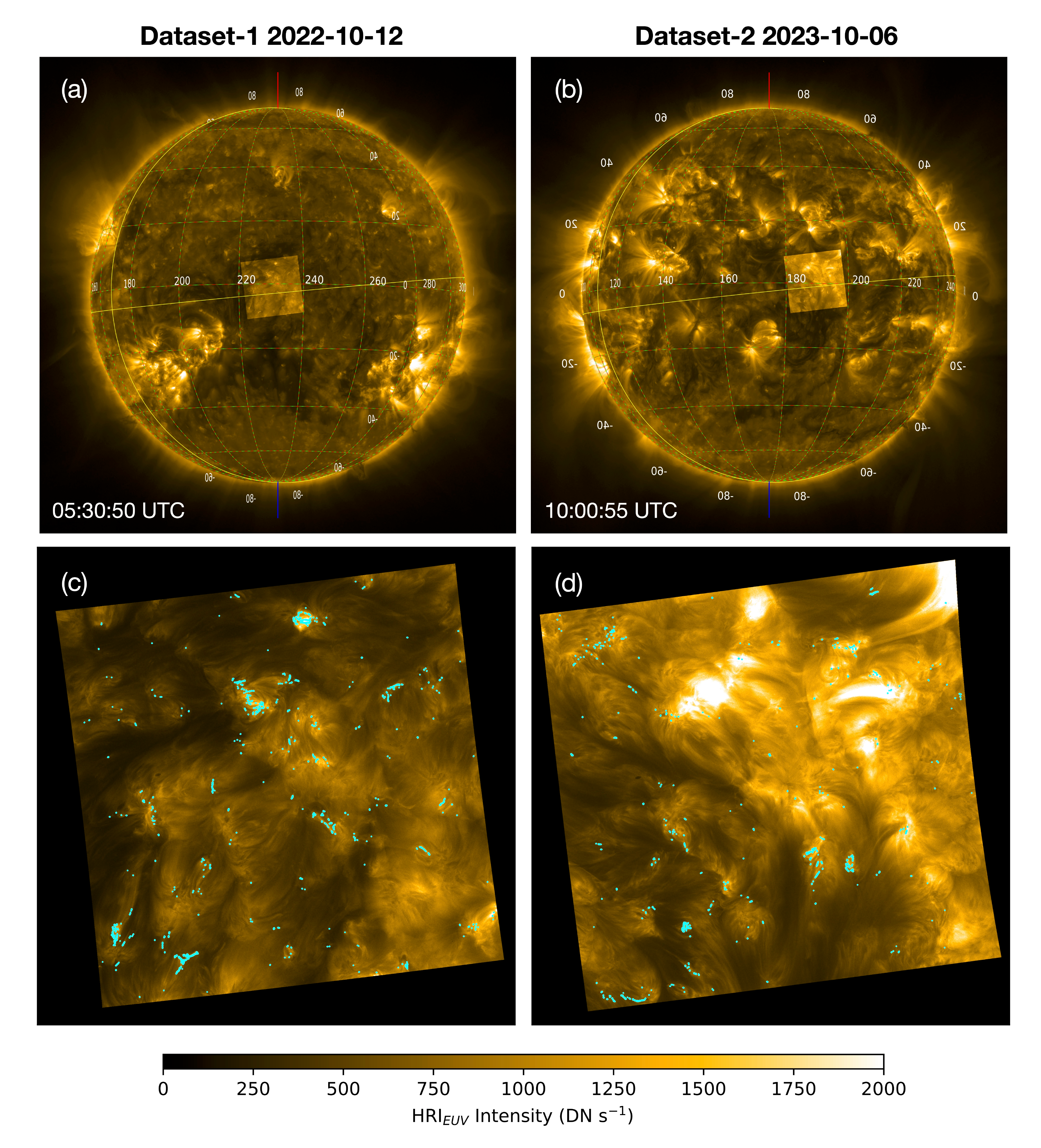}
 	
 	\caption{Representative images of the two EUI quiet-Sun observations studied in this article. Panels (a) and (b) show the visualisation from JHelioviewer \citep{2017Muller}, where the EUI/FSI 17.4\,nm images overlaid with the near-simultaneous HRI\textsubscript{EUV} images are shown. The time mentioned in UTC corresponds to the time of observation at Solar Orbiter. Carrington coordinates are shown, the crossing of the two yellow great circles indicates the antipode of the Earth/SDO sub-solar point. The red and blue markers indicate the Solar north and south poles. Panels (c) and (d) show the HRI\textsubscript{EUV} images that cover a FoV of approximately 20 degrees in longitude and latitude, corresponding to 215\,Mm$\times$215\,Mm on the surface of the Sun for both the datasets. The detected EUV brightenings at the specific instance are marked by cyan color over the respective FoVs for illustration. The online available \href{https://cloud-as.oma.be/index.php/s/xt9ZofwYd7NgKbZ}{movie1} and  \href{https://cloud-as.oma.be/index.php/s/ib29scJX3JackmB}{movie2} show the detected EUV brightenings over the full FoV for the entire duration of the two datasets.}
 	
 	\label{fig1}

 \end{figure*}

 We use the level-2 calibrated release-6 data of EUI/HRI\textsubscript{EUV} \citep{euidatarelease6} available at the EUI website\footnote{\url{https://www.sidc.be/EUI/data/}} and Solar Orbiter Data Archive (SOAR\footnote{\url{https://soar.esac.esa.int/soar/}}). The release-6 data of EUI has considerable improvements in terms of pointing information and provides significantly stable data sequences. The images from dataset-1 were also employed by \citet{Chitta2023} to study properties of quiet-Sun coronal loops and their relation to the underlying surface magnetic field distribution. The duration of dataset-1 is almost 45\,minutes and that of dataset-2 is 60\,minutes. The exposure time for all the images in the two datasets studied here is 1.65 seconds. With a pixel scale of 105\,km, the field-of-view (FoV) of 2048$\times$2048 HRI\textsubscript{EUV} pixels corresponds to 215\,Mm$\times$215\,Mm on the surface of the Sun for both of the datasets. An example of the HRI\textsubscript{EUV} FoV of the two datasets is shown in Fig. \ref{fig1} with context images of the full disk of the Sun as captured by the Full Sun Imager (FSI aboard EUI). During both sets of observations, the Solar Orbiter's orbit was at an inclination of about 4 degrees, towards the solar south, with respect to Earth's orbital plane. The separation angle of the Solar Orbiter with respect to the Sun-Earth line was approximately 120 degrees during these observations, and thus no coordinated observations were possible in these cases with the Earth-aligned observatories.

 
 Note that some portion at the top-right corner of the HRI\textsubscript{EUV} FoV of dataset-2 is covered by coronal loops which belong to a non-flaring active region present in the vicinity of the FoV (see Fig. \ref{fig1} (b) and (d)). As these active region coronal loops cover only a small portion of the FoV and are stable throughout the duration of the data-sequence, we consider dataset-2 as mostly quiet-Sun. Each of the two quiet-Sun datasets used here covers almost 0.75\% of the solar surface and contains the observation sequences for long durations of almost 45--60 minutes. The quiet-Sun spanned by these observations thus can be considered as typical quiet-Sun present over the solar surface at any given instance. Subsequently, the properties of EUV brightenings studied here should be treated as that for the average population of EUV brightenings present over quiet regions of the Sun.




\section{Analysis and Results}

\subsection{Detection of EUV brightenings}

\label{sect3a}

\begin{figure*}[htbp]
	\centering
	
	\includegraphics[width=\hsize]{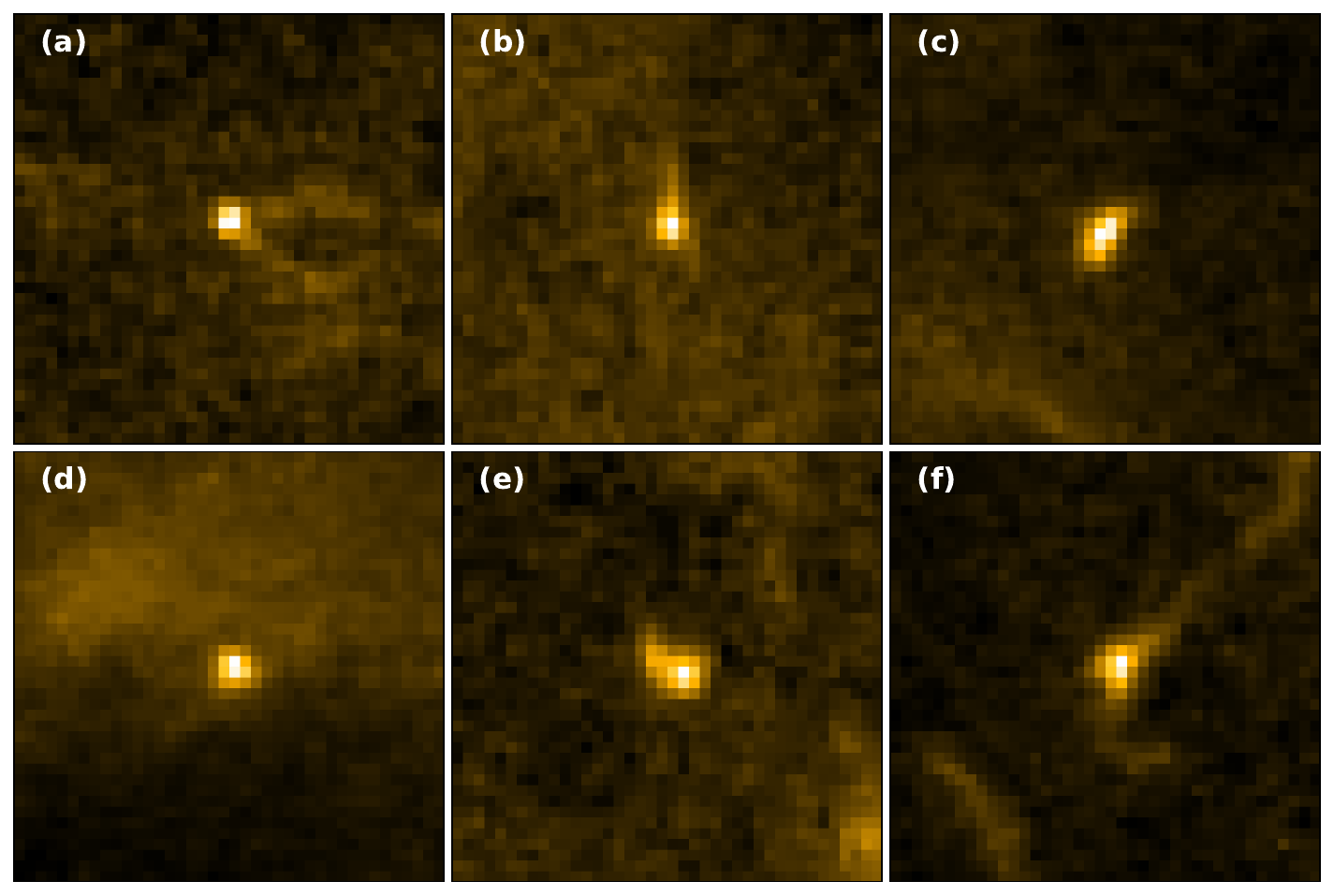}
	
	\caption{Illustrative examples of a close view of six EUV brightenings. The FoV of each panel is 4\,Mm$\times$4\,Mm.}
	
	\label{fig2}

\end{figure*}

 We detect EUV brightenings in both the datasets by using the algorithm employed in \citet{2021Berghmans}. This wavelet-based automated detection scheme separates the small space-time transient events from random intensity fluctuations by determining their statistical significance against the noise present in the data. The relevant details about the detection scheme are provided in Appendix \ref{ApnA}. The detection algorithm reveals the EUV brightenings with their ubiquitous presence across the FoV. We detect a total of 79089 EUV brightenings in dataset-1 and 98649 EUV brightenings in dataset-2. Figure \ref{fig1} (c) and (d) show the detected EUV brightenings at a particular instance over the FoV of the two datasets. We observe that there is tendency for the EUV brightenings to appear in groups (see online available \href{https://cloud-as.oma.be/index.php/s/xt9ZofwYd7NgKbZ}{movie1} and  \href{https://cloud-as.oma.be/index.php/s/ib29scJX3JackmB}{movie2}). Figure \ref{fig2} shows six illustrative examples of a close view of the EUV brightenings, additional examples being provided in Appendix \ref{ApnB}. The smallest EUV brightenings detected have an area of 0.01\,Mm$^{2}$ and the shortest lived EUV brightenings have a lifetime of 3\,s. These minimum values of the detected size and lifetime of EUV brightenings represent the image scale and cadence limit of the observations. Note that the above quoted values do not include the individual events of one spatial-pixel which exist only in a single time-frame. However, the one spatial-pixel events that last for several time-frames and one time-frame events that consist of several spatial-pixels are considered in the analysis. Such events respectively constitute 2\% and 51\% of the total number of EUV brightenings studied here. Thus we are limited more by the temporal cadence than the spatial resolution of the observations in detecting the small-scale events that evolve at much shorter time scales.


 \subsection{Spatial distribution of EUV brightenings}
 
 \label{sect3b}
 
 The average value of the number of EUV brightenings present in every frame of dataset-1 is found to be about 390 and that of dataset-2 is about 350. Their spatial area coverage over the respective FoV is estimated to have an average value of 0.11\% for dataset-1 and 0.08\% for dataset-2. Despite their ubiquitous and pervasive nature, EUV brightenings are observed to avoid the core of the medium to large scale coronal loops present in the FoV. In the close neighbourhood of such structures, the EUV brightenings are preferably present near their foot-points. As mentioned above, the number of EUV brightenings per frame and their spatial area coverage in dataset-2 is slightly less than in dataset-1. This is due to the presence of loop structures in the dataset-2 that cover some portion of the top right corner of the FoV (see Fig. \ref{fig1} (b) and (d)). These coronal loops, as mentioned in Sect. \ref{sect2}, are part of the active region present in the vicinity of the observed FoV.

 With the total number of EUV brightenings detected in each dataset and dividing it by the spatial area of the FoV and duration of respective dataset, we find the occurrence rate of the EUV brightenings from dataset-1 to be 6.3$\times$10$^{-16}$\,m$^{-2}$\,s$^{-1}$ and that from dataset-2 to be 5.9$\times$10$^{-16}$\,m$^{-2}$\,s$^{-1}$. This gives an estimated average occurrence rate of EUV brightenings over the whole surface area of the Sun of approximately 3600 per second. These values are more than one order of magnitude higher than the EUV brightenings reported in \citet{2021Berghmans} and \citet{2024Nelson} and the EUV bursts reported in \citet{2021Chitta}. Note that \citet{2021Berghmans} and \citet{2024Nelson} have used the same detection algorithm as in this sudy, while the detection scheme used in  \citet{2021Chitta} was compeletly different. The increase in the number of detected EUV transient brightenings is thus primarily due to the extremely high resolution of HRI\textsubscript{EUV} at Solar Orbiter perihelion in comparison to other observations (as described in Sect. \ref{sect1} and \ref{sect2}). However, if the two datasets studied here represent unusual quiet-Sun regions, that display an excess of enhanced small-scale transient activity in form of EUV brightenings, the ubiquitous and pervasive presence of EUV brightenings may not be the average nature of a common quite-Sun region. In such a case the high occurence rate of EUV brightenings detected here could be the combined effect of higher resolution and an enhanced small-scale activity of these exceptional quiet-Sun regions. Neverthless, the EUV brightenings detected and studied here are the most prevalent, localised and smallest-scale quiet-Sun EUV transient brightenings detected in the solar corona to date.

\subsection{Morphological and photometrical properties}

\label{sect3c}

\begin{figure}[htbp]
 	\centering
 	\includegraphics[width=\hsize]{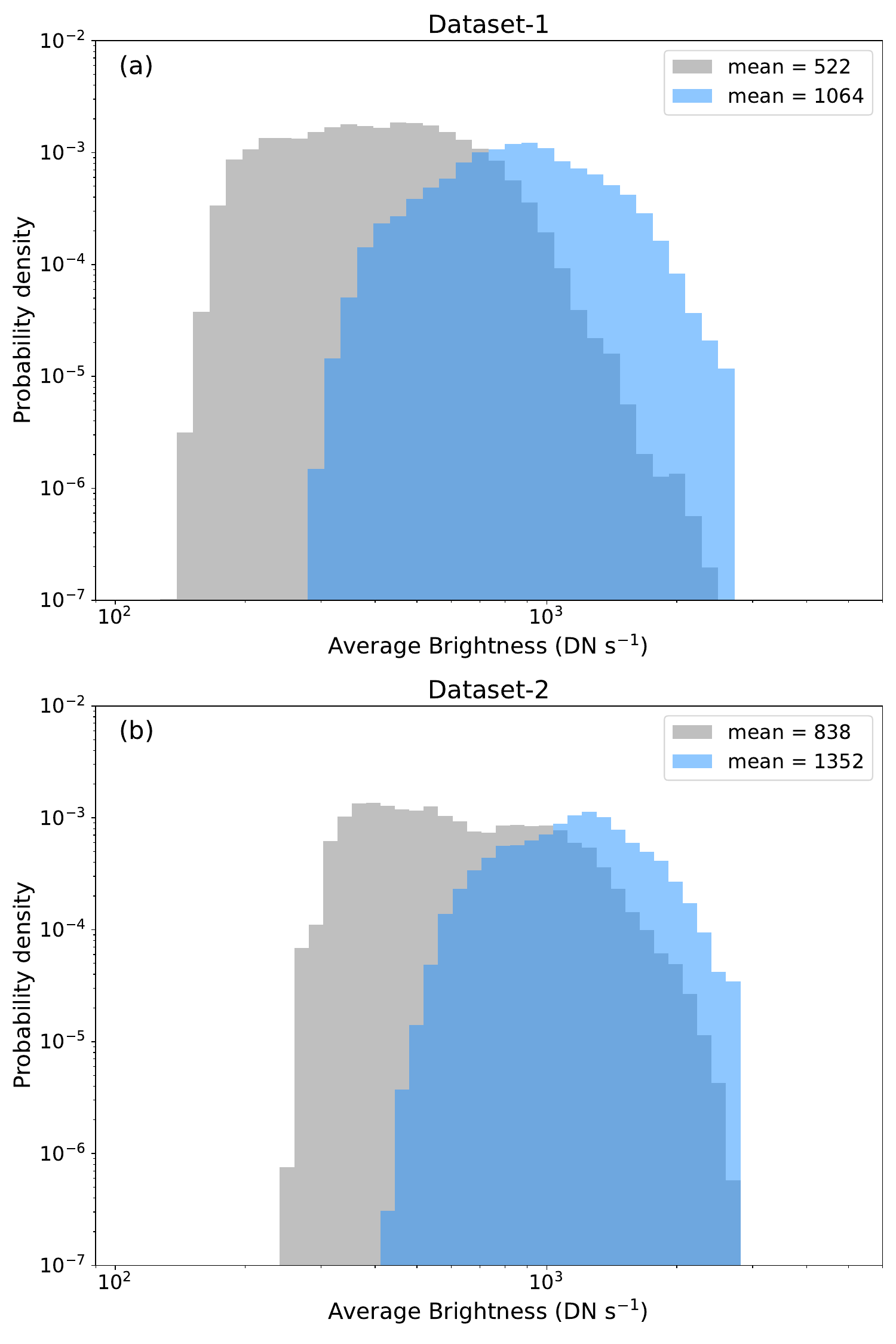}
 	\caption{The distributions of the average brightness of EUV brightenings detected in the two datasets are shown in blue color in each panel. Over-plotted are the respective distributions of pixel brightness of temporally averaged FoV in grey color. The overlapping region of the two distributions is shown with the dark blue color. The axes of both plots are shown with a log\textsubscript{10} scale.}
 	
 	\label{fig3} 
 \end{figure}

 We estimate the morphological properties (surface area, lifetime, volume and aspect ratio) and photometrical properties (total brightness, average brightness and peak brightness) of all the EUV brightenings detected in both the datasets. Appendix \ref{ApnA} provides the definition for the above mentioned properties of the EUV brightenings. Figure \ref{fig3} shows, for both the datasets, the distribution of the average brightness of the EUV brightenings in comparison to the pixel brightness values of the temporally averaged FoV. The distributions of the average brightness of EUV brightenings fall along the higher values of the respective distributions for the pixel brightness of the average FoV. As quoted in Fig. \ref{fig3}, the mean values of the distribution for average brightness of the EUV brightenings are about 1.6 to 2.0 times larger than the mean values of the distributions for the pixel brightness values of the average FoV. The temporally averaged FoV can be considered to represent the local background conditions. This indicates that the average member of the population of EUV brightenings detected in our observations has about a 50\% brightness enhancement over the background quiet-Sun emission.

 
\begin{figure*}[htbp]
 	\centering
 	\includegraphics[width=\hsize]{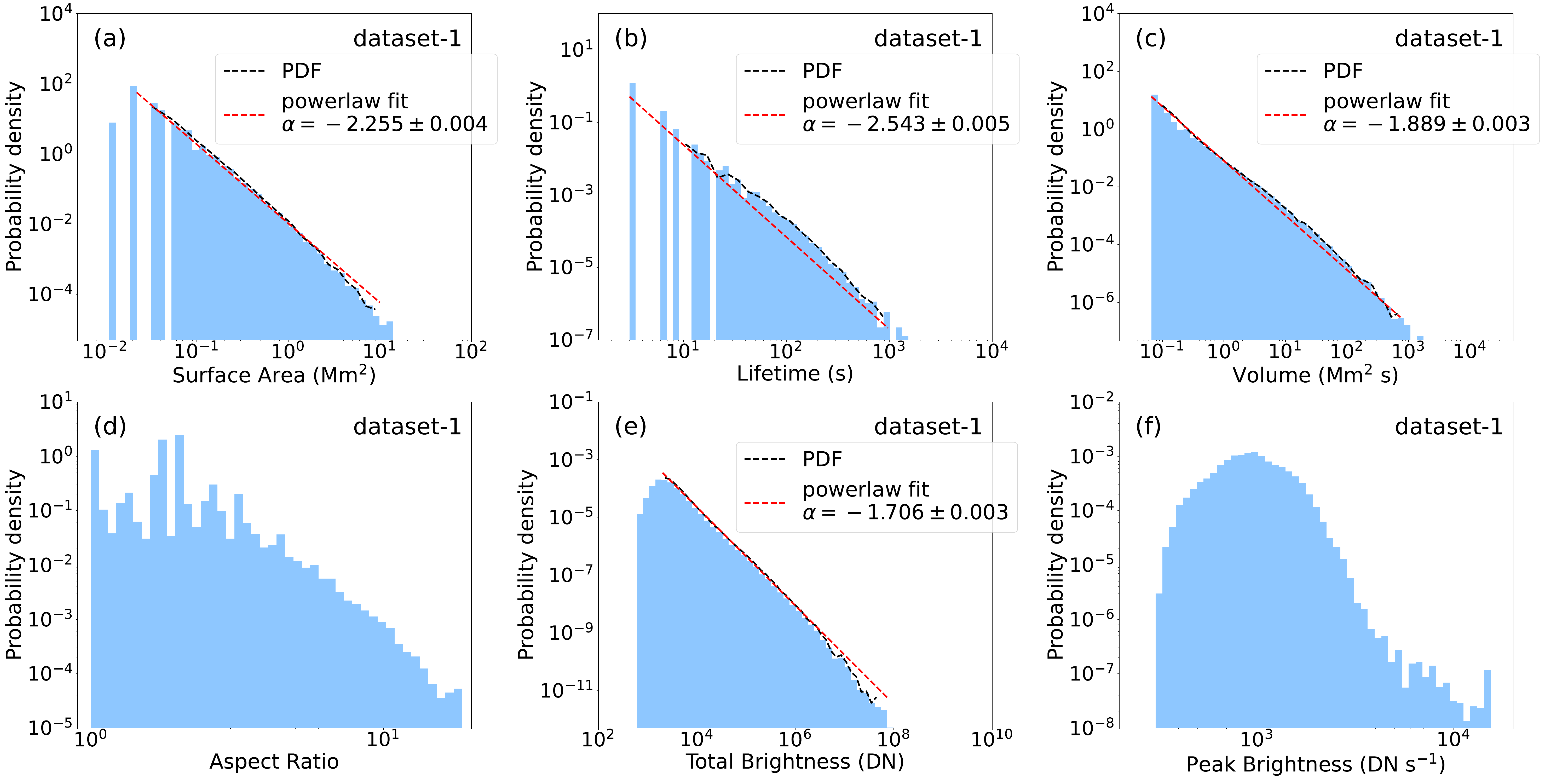}
 	\caption{Probability density distributions of the properties of EUV brightenings for dataset-1. The probability density function (PDF) and power-law fit are shown for the distributions of surface area, lifetime, volume and total intensity. The axes of all the plots are shown with a log\textsubscript{10} scale. The corresponding figure for dataset-2 is shown in Appendix \ref{ApnC}.}

 	\label{fig4}

 \end{figure*}

\begin{figure*}[htbp]
 	\centering
 	\includegraphics[width=\hsize]{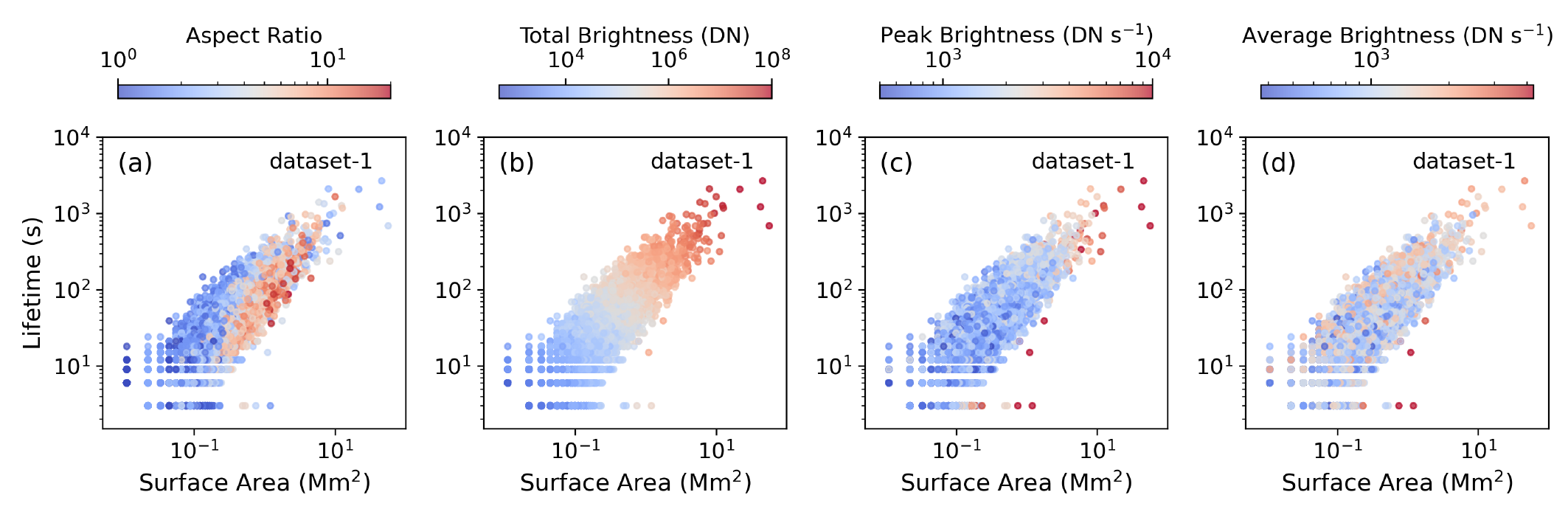}
 	\caption{Relations between the different properties of EUV brightenings for dataset-1. The scatter-plot between surface area and lifetime is shown, with their interrelation among other morphological and photometrical properties in different panels. The axes of all the plots are shown with a log\textsubscript{10} scale. The corresponding figure for dataset-2 is shown in Appendix \ref{ApnC}.}
 	
 	\label{fig5} 
 \end{figure*}

 The distribution of properties of EUV brightenings are studied further with the help of histograms and scatter-plots that are respectively shown, in Fig. \ref{fig4} and Fig. \ref{fig5} for dataset-1, and in Fig. \ref{figC1} and Fig. \ref{figC2} for dataset-2 (Appendix \ref{ApnC}). Noticeably, the respective distributions for the two datasets show very similar trends. The detected EUV brightenings have surface areas in the range of 0.01\,Mm$^{2}$ to 50\,Mm$^{2}$  and lifetimes from 3\,seconds to 40\,minutes. The distribution of the properties of EUV brightenings in Fig. \ref{fig4} shows that, within the detected population of EUV brightenings, their surface area and lifetime vary by almost three orders of magnitude. We perform a maximum-likelihood power-law fit \citep{Clauset2009, Alstott2014} to the  probability density functions estimated from the histograms of surface area, lifetime, volume and total brightness. The calculated power-law indices are quoted in the respective panels of Fig. \ref{fig4} for dataset-1, and similarly in Fig. \ref{figC1} of appendix \ref{ApnC} for dataset-2. The obtained values of power-law indices for the distribution of size and lifetime are similar to those mentioned in \citet{1998Berghmans} for quiet-Sun EUV transient brightenings observed by EIT (Extreme-ultraviolet Imaging Telescope, \citealt{1995Delab}) aboard SOHO mission. This indicates that the small-scale EUV brigtenings detected by different resolution instruments may follow the similar scaling laws and are possibly part of the same nanoflare family. The histogram of aspect ratio in Fig. \ref{fig4} reveals that the events with almost symmetric shape (aspect ratio\,$<$\,3.0) are more numerous than those with assymetric shape (aspect ratio\,$>$\,3.0). Given the power-law behaviour of the distribution of surface area, lifetime and volume, this indicates that the smaller, short lived and events with symmetric morphology dominate in the detected population of EUV brightenings.

 
 The scatter-plots in Fig. \ref{fig5} for dataset-1 show the interrelation between different properties of EUV brightenings. The different panels in Fig. \ref{fig5} show the scatter-plot between surface area and lifetime and their association with aspect ratio in panel (a), total brightness in panel (b), peak brightness in panel (c) and average brightness panel (d). The corresponding scatter-plots for dataset-2 are shown in Fig. \ref{figC2} of appendix \ref{ApnC}. The value of Spearman's correlation coefficient between surface area and lifetime is found to be 0.71 for both the datasets. The scatter-plot shown in Fig. \ref{fig5}\,(a) hints towards two populations of the EUV brightenings showing distinct behaviour with respect to aspect ratio depending on whether their surface area is larger or smaller than 0.2 Mm$^{2}$. In the total EUV brightenings detected in each dataset, almost 93\% of the events have surface area less than 0.2 Mm$^{2}$. The lifetimes of the EUV brightenings in the population with surface area smaller than 0.2 Mm$^{2}$ lie between 3\,s and 2\,min. Among this population group 91\% of EUV brightenings are of almost symmetric shape with aspect ratio less than 3.0. While among the population group with surface area larger than 0.2 Mm$^{2}$, 55\% of the EUV brightenings are of asymmetric shape with aspect ratio more than 3.0. Among these two population groups, there is no distinction observed in terms of their spatial distribution across the FoV. This illustrates that, though randomly spread across the FoV, the smaller and short-lived EUV brightenings are dominated by almost circular shape events, and larger and longer-lived events are mostly elongated in shape.
 
 The scatter-plots shown in Figs. \ref{fig5}\,(b),\,(c)\,and\,(d) show that the morphological properties of EUV brightenings have positive association with their photometrical properties. As shown in Fig. \ref{fig5}\,(b) there is a strong positive association of total brightness with respect to the variation of surface area and lifetime. This is primarily due to the intrinsic definition of total brightness of the event, that is the integrated brightness of the event over its area and duration. Such a positive association of surface area and lifetime is also present for the case of peak brightness (Fig. \ref{fig5}\,(c)) though the trend is not as strong as that for the case of total brightness. For the case of average brightness (Fig. \ref{fig5}\,(d)) the trend is comparatively weaker. Such associations between surface area, lifetime, aspect ratio and brightness of the EUV brightenings show that there is a tendency for spatially smaller, symmetrical and shorter-lived EUV brightenings to have lower values of brightness, and larger, asymmetrical and longer-lived EUV brightenings to have higher values of brightness. The statistical distribution and interrelation among different properties of EUV brightenings presented in this section show that the population group of small events (0.01\,Mm$^{2}$\,$<$\,surface area\,$<$\,0.2\,Mm$^{2}$), that are mostly short-lived (3\,s\,$<$\,lifetime\,$<$\,2\,min), with almost circular shape and they are not very bright in comparison to others, represent the dominant group within the detected population of the EUV brightenings. Such a population of fine-scale EUV brightenings has become acessible by the virtue of high spatial and temporal resolution of HRI\textsubscript{EUV} with its high radiometric sensitivity.




\section{Discussion}

\label{sect4}

 Previous studies of small-scale coronal transients indicate that nanoflare heating might not be sufficient to account for the coronal energy losses outside the active regions of the Sun \citep{2000bAschwanden, 2002Benz, 2016Joulin, 2021Chitta, 2022Phurkart}, primarily due to the low occurrence rate of such low energy events in the quiet-Sun. With the high spatial and temporal resolution observations of HRI\textsubscript{EUV} presented in this study, an increasingly high occurrence rate of the EUV brightenings is observed at finer spatial and temporal scales. If these events reach coronal temperatures of 1\,MK or more, they can be possible candidates for quiet-Sun coronal heating. On the other hand, the high occurrence rate, and ubiquitous and pervasive nature of EUV brightenings observed in the two datasets studied here may not represent the average behaviour of quiet-Sun corona in general. For instance, \citet{Gorman2023} have shown specific cases of large sections of the quiet-Sun corona, on supergranular scales, that lack clear signatures of transient EUV brightening-type events, giving rise to the coronal appearance that is very diffuse.

 \citet{2021Berghmans} have derived temperatures of about 1\,MK for most of the EUV brightenings observed by HRI\textsubscript{EUV}, while their physical heights have been found to be in the transition region to lower corona \citep{2021Zhukov}. By studying a sample of the EUV brightenings \citet{Huang2023} and \citet{2023Dolliou, 2024Dolliou} have indicated that many of these events may have temperatures below 1\,MK and are chromospheric or transition region events. \citet{2023Nelson} have also shown that many of the EUV brightenings display a discernible transition region response. However due to the limited resolution of the supplementary imaging and spectral observations used with HRI\textsubscript{EUV} in the above mentioned studies, significant conclusions cannot be made about the smallest-scale EUV brightenings observed by HRI\textsubscript{EUV}. While it is important to note that by using a 3D self-consistent quiet-Sun model of resolution comparable to HRI\textsubscript{EUV}, \citet{2025Chen} have indicated that the EUV brightenings observed by HRI\textsubscript{EUV} may comprise a mix populations of events with some of the events reaching coronal temperatures of 1\,MK and above, while others remain at transition region temperatures.

 Estimate of temperature and thermal energy flux for the EUV brightenings is crucial to address their role in coronal heating. The HRI\textsubscript{EUV} passband centered at 17.4\,nm has an effective width of 0.5\,nm \citep{2023Gissot}. With the peak temperature response of the HRI\textsubscript{EUV} passband close to 1\,MK, the HRI\textsubscript{EUV} observations can have contributions from solar plasma with temperatures in the range of 0.3\,MK to 1.2\,MK \citep{Shestov25}. Thus multi-wavelength imaging and spectral observations with high spatial, temporal and wavelength resolution are crucial to have better understanding about the temperature and density of the smallest-scale HRI\textsubscript{EUV} brightenings. The joint observations of EUI/HRI\textsubscript{EUV} with upcoming misions in near future, such as Solar--C/EUVST (EUV High-Throughput Spectroscopic Telescope, \citealt{2019Shimizu}) and MUSE (Multi-slit Solar Explorer, \citealt{2022Pontieu}), will be of utmost importance in this context.




\section{Conclusions}
\label{sect5}

 In this article, we have presented the highest-resolution observations of the quiet solar corona available to date. The HRI\textsubscript{EUV} telescope of EUI onboard Solar Orbiter has obtained such observations during the perihelia of Solar Orbiter. Using two such sets of observations we have detected the smallest, shortest-lived and most prevelant EUV transient brightenings reported till now, with sizes down to 0.01\,Mm$^{2}$ and lifetimes of 3\,s. We find that the EUV brightenings are ubiquitous, with 79089 EUV brightenings present in dataset-1 and 98649 in dataset-2. Their spatial area coverage is about 0.1\% with a high occurrence rate of about  6.1$\times$10$^{-16}$\,m$^{-2}$\,s$^{-1}$ over quiet regions of the Sun. This gives an average occurrence rate over the whole surface area of the Sun of approximately 3600 EUV brightenings per second. This occurence rate is more than one order of magnitude greater than that for the previously reported values for the similar events \citep{2021Chitta, 2021Berghmans, 2024Nelson}. This is primarily by the virtue of high spatial and temporal resolution observations of HRI\textsubscript{EUV} obtained at the closest perhelia of the Solar Orbiter. Given that the observations used in this study have the best possible spatial resolution achievable by HRI\textsubscript{EUV}, higher-cadence observations of the quiet Sun, while maintaining the highest possible spatial resolution, may reveal an even higher occurrence rate of EUV brightenings. Such EUI/HRI\textsubscript{EUV} observations are planned for the Solar Orbiter's perihelion in the near future. However, great caution must be exercised regarding the exposure time of faster-cadence observations to maintain sufficient signal levels, particularly when observing the low-emission quiet regions of the Sun.
 
 We have studied the statistical distribution and interrelation among different morphological and photometrical properties of EUV brightenings. We find power-law behaviour in the distribution of the surface area and lifetime of the EUV brightenings. A positive correlation is observed to be present between these properties of the brightenings. We observe that among the detected EUV brightenings spatially smaller and shorter-lived events predominantly have almost symmetric/circular shape. Whilst the larger and longer duration events mostly have asymmetric/elongated morphology. There also exists a positive association of brightness of the EUV brightenings with their surface area and lifetimes. We observe a general trend of smaller and shorter-lived events to have lower values of brightness, and larger and longer-lived to have higher values of brightness. 
 
 We find that EUV brightenings with surface area smaller than 0.2\,Mm$^{2}$  and lifetimes less than 2\,min represent the dominant population group within the detected EUV brightenings. Thus the high spatial and temporal resolution of HRI\textsubscript{EUV} with its high radiometric sensitivity is indispensable to study the large ensemble of smallest-scale EUV brightenings in detail. However, understanding of the thermal properties of the HRI\textsubscript{EUV} transient brightenings remains elusive due to the current absence of high-resolution multi-wavelength imaging and spectral observations. Such high resolution observations coordinated with HRI\textsubscript{EUV} which can simultaneously map transition region and corona are necessary in this context. Future solar missions such as MUSE and Solar--C/EUVST, in combination with Solar Orbiter/EUI observations, will provide unique opportunities to gain better insights about the finest scale events in the solar corona.


\begin{acknowledgements}
Authors thank anonymous referee for constructive comments. Solar Orbiter is a space mission of international collaboration between ESA and NASA, operated by ESA. The EUI instrument was built by CSL, IAS, MPS, MSSL/UCL, PMOD/WRC, ROB, LCF/IO with funding from the Belgian Federal Science Policy Office (BELSPO/PRODEX PEA 4000106864, 4000112292 and 4000134088); the Centre National d$’$Etudes Spatiales (CNES); the UK Space Agency (UKSA); the Bundesministerium f\"{u}r Wirtschaft und Energie (BMWi) through the Deutsches Zentrum f\"{u}r Luftund Raumfahrt (DLR); and the Swiss Space Office (SSO). N.N. acknowledges funding from the Belgian Federal Science Policy Office (BELSPO) contract B2/223/P1/CLOSE-UP. C.V. and L.D. thanks BELSPO for the provision of financial support in the framework of the PRODEX Programme of the European Space Agency (ESA) under contract numbers 4000143743, 4000134088 and 4000136424. L.P.C. gratefully acknowledges funding by the European Union (ERC, ORIGIN, 101039844). Views and opinions expressed are however those of the author(s) only and do not necessarily reflect those of the European Union or the European Research Council. Neither the European Union nor the granting authority can be held responsible for them. This research was supported by the International Space Science Institute (ISSI) in Bern, through ISSI International Team project \#23-586 (Novel Insights Into Bursts, Bombs, and Brightenings in the Solar Atmosphere from Solar Orbiter). Authors thank Dr. Susanna Parenti and Prof. Louise Harra for helpful discussions. \textit{JHelioviewer} software was used in this research to visualise EUI observations. \textit{SSWIDL} and various \textit{Python} packages including \textit{sunpy} and \textit{astropy} were used for the data analysis. This work has used NASA's \textit{Astrophysics Data System}.
\end{acknowledgements}

\bibliographystyle{aa} 
\bibliography{euv_bright_rob} 


\begin{appendix}

 \section{Detection scheme for the EUV brightenings}
    
 \label{ApnA}

  As mentioned in Sect. \ref{sect3a} we use the wavelet-based automated detection algorithm similar to that employed in \citet{2021Berghmans}, to isolate the transient EUV brightenings occurring at small scales in the observations. We separate the small features of interest spatially by determining their statistical significance over the noise present in the data and track them in time through the observation sequence. The HRI\textsubscript{EUV} level-2 images are remapped to Carrington coordinates, with 105\,km pitch, before subjecting them to the detection algorithm. The HRI\textsubscript{EUV} resolution is thus kept preserved while performing the coordinate transformation.

  In the noise model we consider the photon shot noise and read-out noise which are the dominant sources of noise present in the HRI\textsubscript{EUV} observations \citep{2020Rochus, 2023Gissot}. The small-scale transient brightenings of interest are detected using the first two scales of a dyadic \textit{`\`{a} trous'} wavelet transform of the images using a \textit{B\textsubscript{3}} spline scaling function \citep{1994Starck, 2002Starck}. Using the treatment of \citet{1995Murtagh} and applying Poisson statistics, wavelet coefficients in the first two scales are considered significant when they are greater than 6 times the root-mean-square amplitude expected from the noise model. Such a threshold puts a rigorous statistical significance criterion on the detection of the small-scale events with almost 99\% confidence level against the dominant sources of noise present in the data. The thresholding of the wavelet coefficients in each image results in a binary cube. The 6-connected voxels in the space and time \textit{(x,y,t)} dimensions of the binary cube are clustered into numbered regions. Each region defines an event \textit{i.e.} an EUV brightening.

   The above detection scheme is applied to both the datasets. The morphological properties (lifetime, surface area, volume and aspect ratio) and photometrical properties (total brightness, average brightness, and peak brightness) are estimated for of all the EUV brightenings detected in both the datasets. The lifetime of an event is its duration in seconds. The surface area of an event is given by the area in Mm$^{2}$ in the image plane, of the total number of pixels of the projection of the event along the temporal axis during its lifetime. The volume is defined as the total number of voxels of the event including its spatial and temporal pixels and is expressed in Mm$^{2}$\,s. The estimate of aspect ratio is obtained by the ratio of major axis to minor axis of the ellipse fitted to the event at the instance of its largest spatial extent. The total brightness of an event, expressed in data-numbers (DN), is calculated as the integrated data value over all the voxels of an event during its lifetime. The average brightness of an event, in DN\,s$^{-1}$, is the value obtained by dividing the total brightness value by the total number of voxels of the event. The peak brightness, expressed in DN\,s$^{-1}$, is the data value of the brightest voxel of the event. The detection results presented in this study are also employed by \citet{Lim25} to study quasi-periodic pulsations in the EUV brightenings.

 \section{Examples of the EUV brightenings}

 \label{ApnB}

\begin{figure*}[htbp]
    \centering
    \includegraphics[width=\hsize]{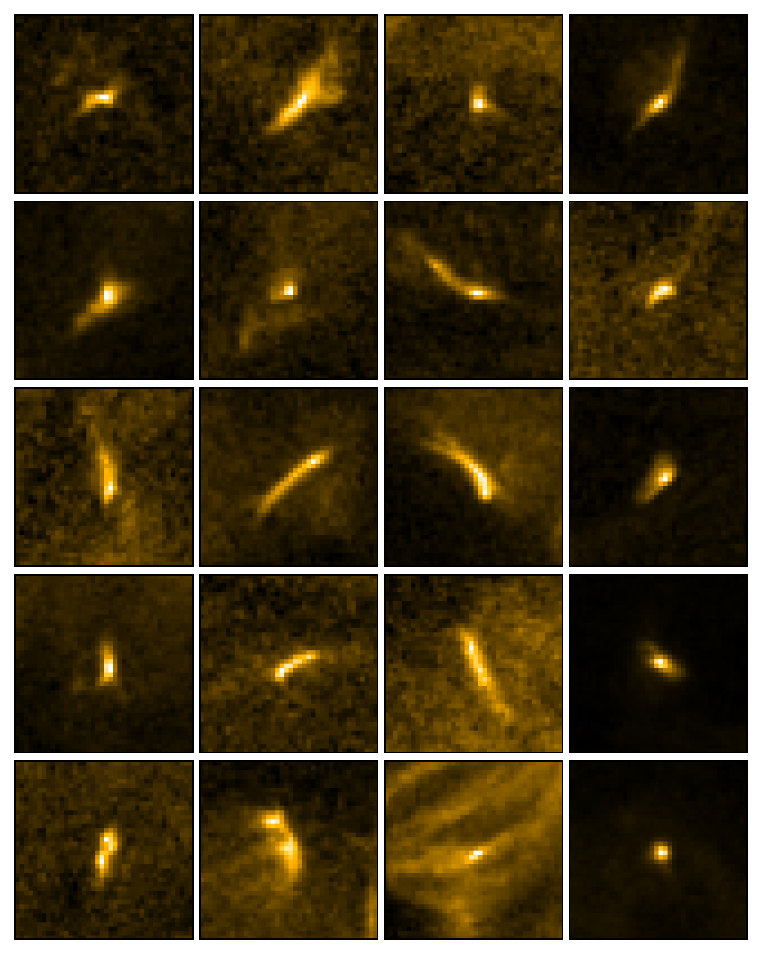}
    \caption{Additional examples of close view of 20 EUV brightenings. The FoV of each panel is 4\,Mm$\times$4\,Mm. See appendix \ref{ApnB}.}
    \label{figB1}

    \end{figure*}

  Figure \ref{figB1} shows additional examples of EUV brightenings.

  \section{Distribution of properties of EUV brightenings for dataset-2}
 
 \label{ApnC}
 
\begin{figure*}[htbp]
 	\centering
 	\includegraphics[width=\hsize]{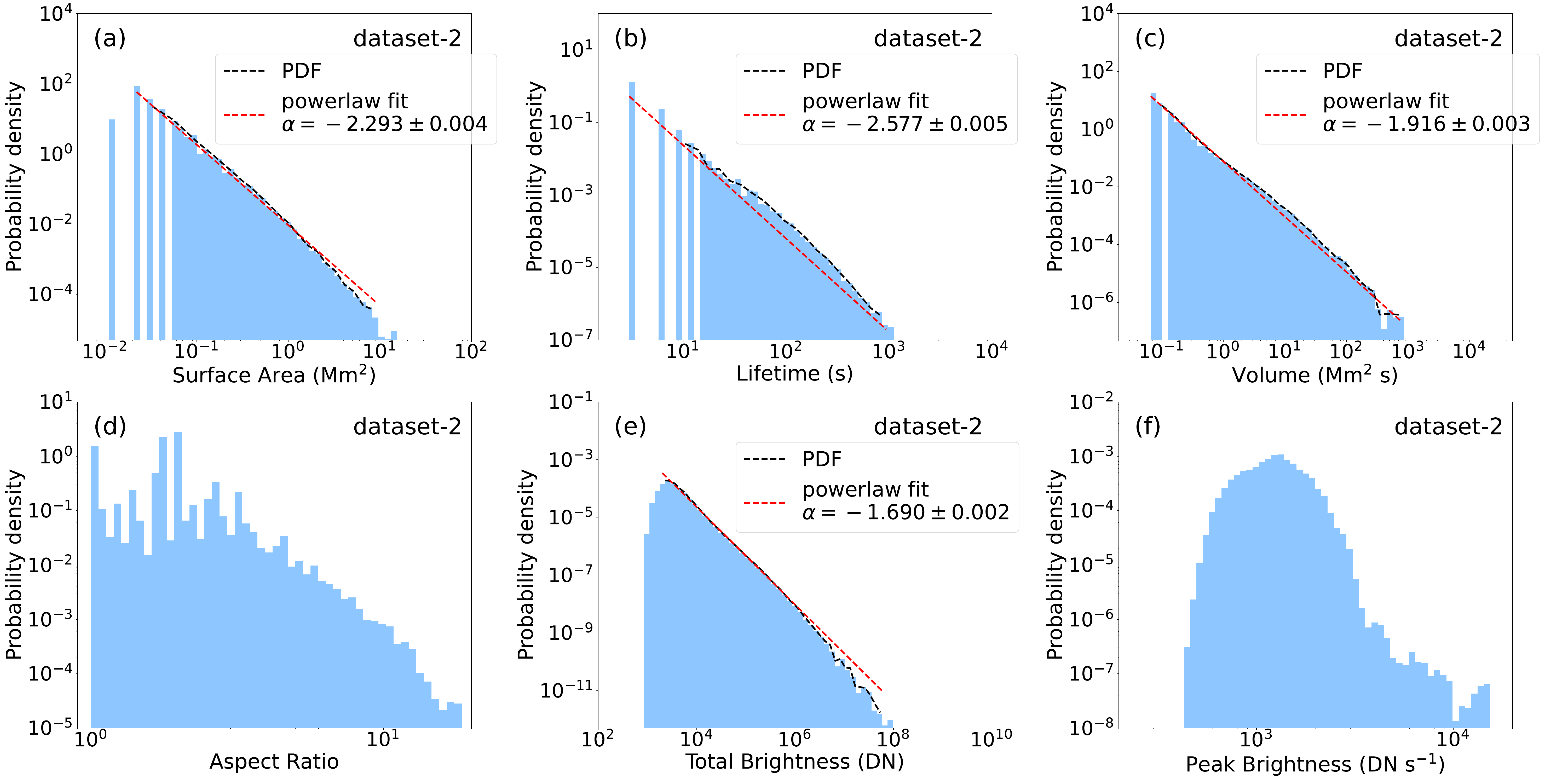}
 	\caption{Same as Fig. \ref{fig4} but for dataset-2. See appendix \ref{ApnC}.}

 	\label{figC1}

 \end{figure*}

\begin{figure*}[htbp]
 	\centering
 	\includegraphics[width=\hsize]{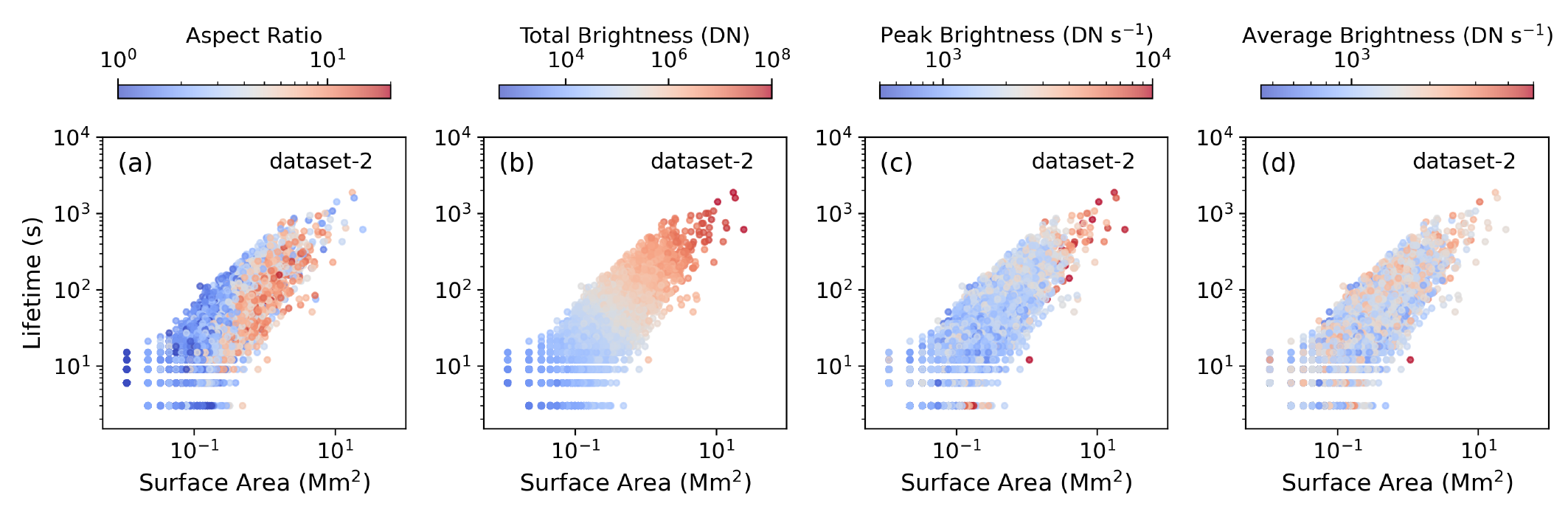}
 	\caption{Same as Fig. \ref{fig5} but for dataset-2. See appendix \ref{ApnC}.}
 	
 	\label{figC2} 
 \end{figure*}

	 Figures \ref{figC1} and \ref{figC2} respectively show the histograms and scatter-plots for the properties of EUV brightenings for dataset-2. The corresponding figures for dataset-1 are shown as figs. \ref{fig4} and \ref{fig5} in Sect. \ref{sect3c}. The trends in the different probability distributions and interrelation among different properties of the EUV brightenings are very similar for both the datasets, that are explained in Sect. \ref{sect3c}.

 
 \end{appendix}
 
\end{document}